\documentclass[11pt,a4paper,abstract=on]{scrartcl}

\usepackage[utf8]{inputenc}
\usepackage[english]{babel}

\usepackage[paper=a4paper,left=25mm,right=25mm,top=25mm,bottom=30mm]{geometry}
\usepackage{amsmath,amssymb,amsthm,mathrsfs,bbm,amsfonts,bm}
\usepackage{color,graphicx,cases,caption,enumerate}
\usepackage{float,multicol,authblk}
\usepackage[dvipsnames]{xcolor}
\usepackage{booktabs}
\usepackage{multirow}
\usepackage{subcaption}
\usepackage{algorithmic}
\usepackage{booktabs}
\usepackage{capt-of}
\usepackage{array}
\usepackage[title]{appendix}
\newcolumntype{C}[1]{>{\centering\let\newline\\\arraybackslash\hspace{0pt}}m{#1}}
\usepackage{natbib}
\usepackage{placeins} 
\usepackage[breaklinks, colorlinks=true, citecolor=black, urlcolor=black, linkcolor=black]{hyperref}

\newcommand{\rawpp}{GHI$^\text{raw}$-PV$^\text{pp}$\ }
\newcommand{\ppraw}{GHI$^\text{pp}$-PV$^\text{raw}$\ }
\newcommand{\pppp}{GHI$^\text{pp}$-PV$^\text{pp}$\ }
\newcommand{\rawraw}{GHI$^\text{raw}$-PV$^\text{raw}$\ }

\title{Improving Model Chain Approaches for Probabilistic Solar Energy Forecasting through Post-processing and Machine Learning}

\author[1]{Nina Horat}
\author[1]{Sina Klerings}
\author[1,2]{Sebastian Lerch}
\affil[1]{Karlsruhe Institute of Technology}
\affil[2]{Heidelberg Institute for Theoretical Studies}

\date{\today}

\begin{document}

\maketitle

\begin{abstract}
\noindent
Weather forecasts from numerical weather prediction models play a central role in solar energy forecasting, where a cascade of physics-based models is used in a model chain approach to convert forecasts of solar irradiance to solar power production, using additional weather variables as auxiliary information.
Ensemble weather forecasts aim to quantify uncertainty in the future development of the weather, and can be used to propagate this uncertainty through the model chain to generate probabilistic solar energy predictions.
However, ensemble prediction systems are known to exhibit systematic errors, and thus require post-processing to obtain accurate and reliable probabilistic forecasts.
The overarching aim of our study is to systematically evaluate different strategies to apply post-processing methods in  model chain approaches: 
Not applying any post-processing at all; post-processing only the irradiance predictions before the conversion; post-processing only the solar power predictions obtained from the model chain; or applying post-processing in both steps.
In a case study based on a benchmark dataset for the Jacumba solar plant in the U.S., we develop statistical and machine learning methods for post-processing ensemble predictions of global horizontal irradiance and solar power generation. 
Further, we propose a neural network-based model for direct solar power forecasting that bypasses the model chain. 
Our results indicate that post-processing substantially improves the solar power generation forecasts, in particular when post-processing is applied to the power predictions. 
The machine learning methods for post-processing yield slightly better probabilistic forecasts, and the direct forecasting approach performs comparable to the post-processing strategies.
\end{abstract}

\section{Introduction}

Reducing greenhouse gas emissions and mitigating climate change requires a rapid transition towards renewable energy \citep{vanderMeerEtAl2018}.
In addition to wind energy, photovoltaic (PV) solar power plays a pivotal role, with decreasing prices and increasing installed capacity in numerous countries. For example, PV power covered 12 percent of the  gross electricity consumption in Germany on average in 2023, and temporarily more than two thirds of the
electricity demand on sunny days \citep{FraunhoferISE}. 
In light of the volatile nature of renewable energy generation and their increasing importance, accurate and reliable forecasts of power generation from those sources are paramount for managing the electrical grid and to balance demand and supply \citep{GottwaltEtAl2016,Appino2018}.
A key development in the energy forecasting literature over the past years has been the transition from single-valued deterministic to probabilistic forecasts \citep{Gneiting2014, Haupt2019, Yang2019a, GneitingEtAl2023} which allow for uncertainty quantification and can be issued in the form of probability distributions, quantiles, or prediction intervals \citep{Lauret2019, Gneiting2023}.

Evidently, weather forecasts from numerical weather prediction (NWP) models are among the most important inputs to models for PV power forecasting. 
A widely used strategy is the conversion of global horizontal irradiance (GHI) forecasts from an NWP system to PV power forecasts via a model chain, potentially using predictions of other meteorological variables as additional inputs \citep{Roberts2017,Mayer2022}. 
The conversion models typically use several meteorological variables such as GHI, temperature, and wind speed as inputs, and require several calculation steps, hence the term ``model chain'', with individual models for the solar position, the separation of beam and diffuse irradiance, the shading loss, the PV performance, and other aspects \citep{Yang2019IJF,WangEtAl2022}.
A variety of possible conversion models or components for individual processes exist and can be utilized to quantify forecast uncertainty by generating an ensemble of model chains \citep{Mayer2021,Mayer2022}.
Further, NWP models are typically run in ensemble mode by generating multiple simulation runs from varying initial conditions and/or changes to the model physics. This process yields a probabilistic forecast in the form of an ensemble, the members of which can be used as inputs to a model chain to generate an ensemble prediction of PV power \citep{WangEtAl2022}.

In the meteorological literature there is broad evidence that NWP ensemble predictions of various weather variables show systematic errors, which require correction to obtain accurate and reliable probabilistic forecasts.  This correction process is called post-processing, for which an overview of common methods and recent developments can be found in \citet{vannitsem2021statistical}. 
Most post-processing methods are statistical or machine learning (ML)-based distributional regression models where calibrated probabilistic forecasts are obtained in the form of parametric probability distributions, quantiles, or corrected ensemble predictions.
One of the most popular post-processing methods is the ensemble model output statistics \citep[EMOS;][]{gneiting2005calibrated} approach, where the forecast takes the form of a parametric distribution, the parameters of which are modeled as functions of summary statistics of the ensemble predictions.
A recent focus of the post-processing literature has been the use of modern ML methods such as random forests \citep{TaillardatEtAl2016} or neural networks \citep[NNs;][]{rasp2018neural}, which allow for incorporating additional meteorological variables beyond the variable of interest as inputs, and have shown substantial improvements in predictive performance over classical statistical approaches such as EMOS, see, e.g., \citet{vannitsem2021statistical} and \citet{Haupt2021} for overviews, and \citet{EUPPBench} for a benchmarking framework.

\begin{figure}
    \centering
    \includegraphics[width=\textwidth]{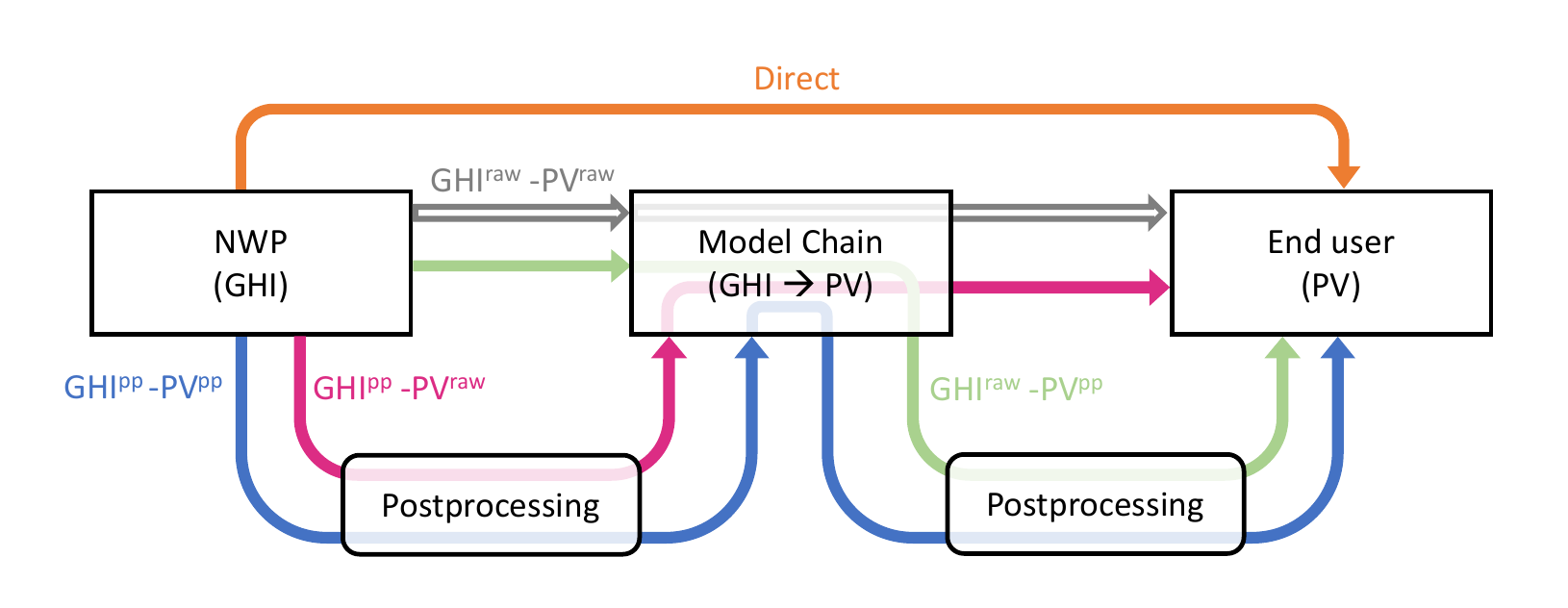}
    \caption{Schematic illustration of the different strategies for applying post-processing methods within a model chain approach for PV power prediction.}
    \label{fig:schematic}
\end{figure}

Statistical and ML-based post-processing methods have also been developed for the purpose of solar energy forecasting, most prominently for post-processing solar irradiance predictions from NWP models \citep[e.g.,][]{Bakker2019,LeGalLaSalle2020,Yang2020EMOS,Yagli2020,Schulz2021,Baran2023,Song2024}.
Since similar post-processing methods can in principle be applied to the PV power predictions obtained as an output of the model chain, this allows for various ways of employing post-processing within probabilistic GHI-to-power conversion approaches utilizing model chains \citep{WangEtAl2022}. 
Following related work on wind energy by \citet{PhippsEtAl2022}, four different strategies are possible: Not applying any post-processing at all and using the raw, unprocessed ensemble predictions obtained as outputs of the model chain (which we will denote by GHI$^\text{raw}$-PV$^\text{raw}$); applying post-processing only to the GHI predictions before the conversion (\ppraw\!\!); applying post-processing only to the PV power forecasts obtained from the model chain conversion (\rawpp\!\!); or applying post-processing in both steps (\pppp\!\!).
Figure \ref{fig:schematic} provides a schematic overview of the different strategies.

The contributions of our work are threefold: First, we systematically evaluate these different strategies to assess the prospects of applying post-processing in model chain approaches, thus extending the work of \citet{PhippsEtAl2022} to solar energy, and related work by \citet{Theocharides2020} to probabilistic forecasts. 
Second, we specifically investigate the use of NN-based post-processing methods for GHI and PV power forecasts.
Third, we compare the different strategies to a NN-based direct probabilistic PV power forecasting model, which uses the meteorological variables as inputs and produces probabilistic forecasts of PV power as its output without applying a model chain for the intermediate conversion step.
Our study is based on a benchmark dataset for solar power forecasting \citep{WangEtAl2022} which comprises weather forecast and PV power observation data for a solar plant in the U.S.

The remainder of the article is organized as follows. Section \ref{sec:data} describes the benchmark dataset and additional data collection and pre-processing steps. Section \ref{sec:methods} introduces the methods used for GHI and PV power post-processing and the models for the conversion from GHI to PV power. Results for the case study are presented in Section \ref{sec:results}, followed by a concluding discussion in Section \ref{sec:conclusions}. Python code with implementations of all models to reproduce the results is available at \url{https://github.com/HoratN/pp-modelchain}.

\section{Data}\label{sec:data}

Our study is based on hourly data for the Jacumba Solar Project in southern California, U.S., covering the years 2017 to 2020.  It comprises four different components, which will be introduced below and the large majority of which is taken from \citet{WangEtAl2022}\footnote{The data is available at \url{https://github.com/wentingwang94/probabilistic-solar-forecasting}.}.
For developing post-processing models, we use the data from July 30, 2017 until end of 2019 as training data and the year 2020 as test data. 

\subsection{Weather predictions}

The weather predictions include ensemble forecasts of GHI from the European Centre for Medium-Range Weather Forecasts (ECMWF), as well as deterministic predictions of additional weather variables from ECMWF's high-resolution (HRES) model. 
The GHI ensemble forecasts have 50 members and are initialized daily at 00 UTC with a lead time of 24 hours.
They contain hourly GHI averages in W~m$^\textnormal{-2}$ and are time-stamped at the full hour marking the end of the averaging period. 
For the conversion from GHI to PV power we use a model chain approach that also takes temperature (in °C) and wind speed (in m~s$^\textnormal{-1}$) as input. Those variables will also be used as additional inputs to the NN-based post-processing models proposed in Section~\ref{sec:methods}.
These forecasts contain instantaneous values at the end of the hour and therefore do not perfectly align with the remaining datasets.
Note that HRES predictions of additional weather variables are available in the original dataset, however, we here follow \citet{WangEtAl2022} and restrict our attention to variables that are directly used within the model chain.

\subsection{GHI observations}

For the purpose of GHI post-processing, we require additional verifying data that can be used as ground truth observations. 
This observational dataset thus is the only part of the data used in our study that is not based on \citet{WangEtAl2022}.
For this purpose, we use satellite-based GHI estimates which we downloaded from the website of the National Solar Radiation Database \citep[NSRDB;][]{sengupta2018}\footnote{\url{https://nsrdb.nrel.gov/}} and which contain hourly irradiance values in W~m$^\textnormal{-2}$ for the location of the Jacumba solar plant (32.62 latitude, -116.13 longitude). 
We presume that the time stamps, e.g., [2017-01-01 00:30], correspond to the middle of the hourly averaging window. 
To ensure consistency with the other datasets (i.e., the weather predictions and the PV power output observations), we therefore adjust the time stamp to the end of the averaging windows, resulting in [2017-01-01 01:00] for the previous example.
Note that for the remainder of the article we refer to the satellite-based GHI estimates (and also the simulated PV power output introduced below) as ``observational'' data since they will be used as ``best estimates'' of the truth.

\subsection{PV power output observations}

Simulated hourly PV power output of the Jacumba solar plant in MW was published by the Lawrence Berkeley National Laboratory (\citet{seel2021pvobsdata}; plant ID: 60947). 
The dataset contains PV power estimates computed with the System Advisor Model (SAM) by the National Renewable Energy Laboratory. 
The first available PV power information are recorded for July~30, 2017 at 00 UTC when the Jacumba solar project was put into operation. 
Since the data contains hourly PV values that are stamped at the beginning of the hour\footnote{as detailed in the user guide available at \url{https://live-etabiblio.pantheonsite.io/sites/default/files/user_guide_for_data_file.pdf}}, we change the time stamp to the end of the averaging window to be consistent with the remaining datasets.

\begin{figure}
\centering
\includegraphics[width=\textwidth]{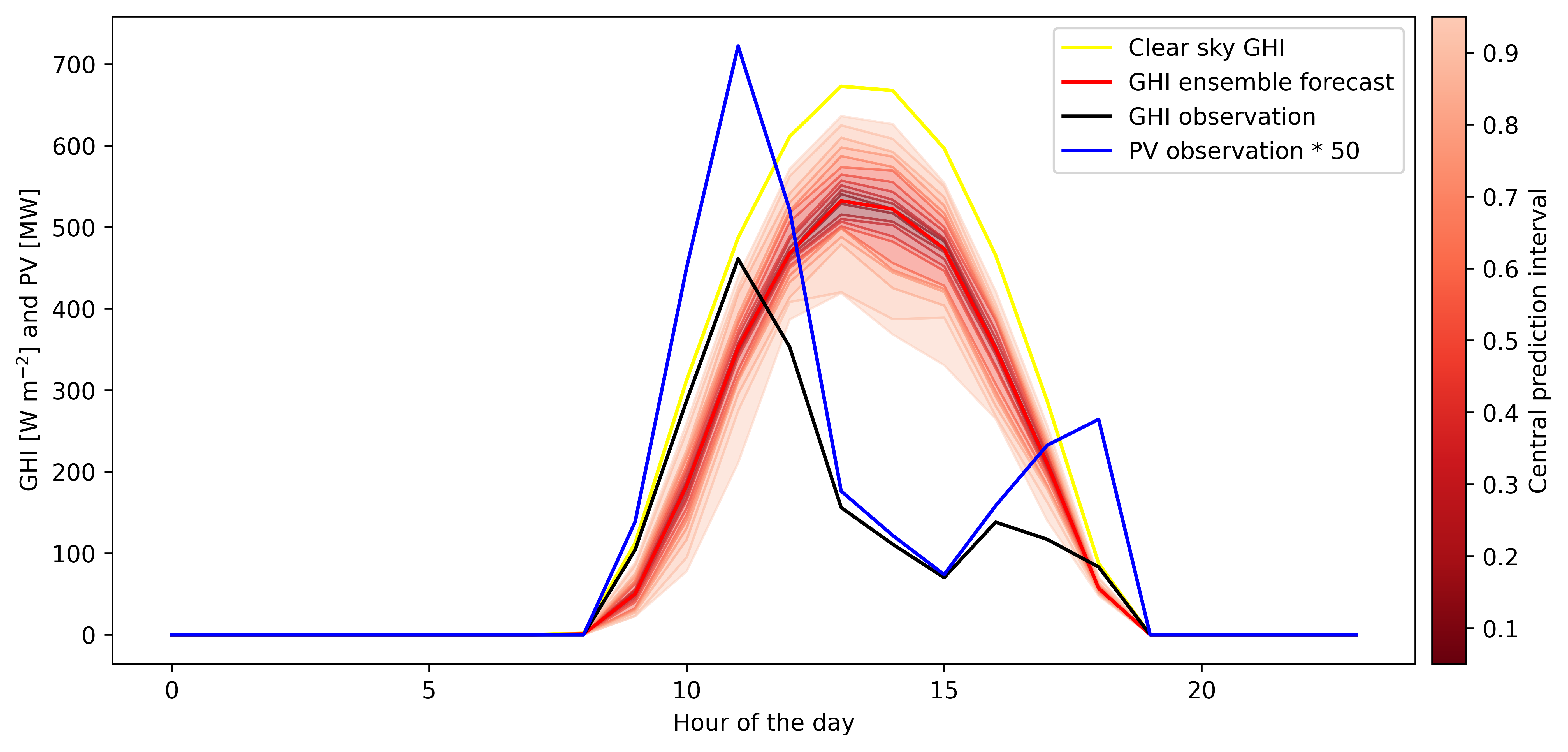}
\caption{Visualization of exemplary GHI ensemble predictions, GHI observations, and PV power observations at the Jacumba solar plant on January 21, 2020. Note that the PV power estimates are scaled by a factor of 50 to allow for a straightforward visual comparison with the GHI datasets, and that the clear-sky GHI values are included here for visualization purposes only.}
\label{fig:data_overview}
\end{figure}

Note that these adjustments of the time stamps deviate from \citet{WangEtAl2022}, however, we found them to be helpful to better align the GHI and PV power observations. 
To illustrate this, Figure \ref{fig:data_overview} shows an exemplary day from all parts of the data for January 21, 2020. 
Due to the time stamp adjustments, the PV and GHI observations are well aligned in time and also match the diurnal cycle present in the ensemble GHI forecasts and the clear sky GHI estimates. 
Note that here and in the remainder of the article, the hour of the day will always refer to the local time at the Jacumba solar plant.

\section{Methods}
\label{sec:methods}

Here, we briefly introduce relevant forecast evaluation methods used in our study, followed by descriptions of the model chain, and the different methods we propose for post-processing the GHI and PV power forecasts, as well as the direct forecasting approach. Results for the different strategies to apply post-processing within the model chain approach will be presented in Section~\ref{sec:results} below.

\subsection{Forecast evaluation}

We here provide a brief overview of the employed evaluation metrics, and refer to \citet{Lauret2019} for a detailed overview specifically tailored to probabilistic solar forecasts, as well as \citet[Section 4]{GneitingEtAl2023}.
It has now been widely accepted that probabilistic forecasts should be as sharp as possible, subject to being calibrated \citep{Gneiting2014}. Calibration refers to the statistical consistency between the forecast distribution and the observation and essentially indicates whether the  observation behaves like a random draw from the forecast distribution. To assess calibration, we use the histograms of the probability integral transform~(PIT) $F(y)$, where $F$ denotes the cumulative distribution function (CDF) of a probabilistic forecast, and $y$ denotes the realizing observation.  
If the probabilistic forecast is calibrated, the PIT histogram should follow a uniform distribution and systematic deviations from uniformity can be used to identify misspecifications of the forecast distributions, see \citet{gneiting2007probabilistic} for details. 
Note that censored forecast distributions with point masses in one or multiple points require adaptations to the calculation of the PIT value to account for the jumps in the forecast CDF. Here, we utilize the randomized PITs proposed in \citet{czado2009predictive}. 
For probabilistic forecasts given in the form of an ensemble, verification rank histograms provide an analogous tool for visual calibration assessment. Thereby, the rank of the realizing observation when pooled with the ensemble predictions should be uniformly distributed. 
We further calculate the coverage and width of central prediction intervals (PIs) with nominal level $\frac{m-1}{m+1}$, where $m$ is the size of the raw ensemble, which is 50 in our case. For a calibrated forecast, the coverage should be close to the nominal value, and the shorter the prediction interval, the sharper the forecast.

Further, proper scoring rules \citep{gneiting2007strictly} enable a simultaneous assessment of calibration and sharpness. The most widely used proper scoring rule in the meteorological literature is the continuous ranked probability score \citep[CRPS;][]{matheson1976scoring},
\begin{equation}
    \label{eq:crps}
    \text{CRPS} (F, y) = \int_{-\infty}^{\infty} \left( F(z) - \mathbbm{1} \lbrace y \leq z \rbrace \right)^2 dz, 
\end{equation}
where $F$ is a forecast CDF with finite first moment, $\mathbbm{1}$ is the indicator function, and $y$ is the observation.
Closed-form analytical expressions of the integral in \eqref{eq:crps} are available for a variety of parametric distributions, including the censored normal distributions we use below, see \citet{JordanEtAl2019} for details.

To evaluate the accuracy of deterministic forecasts  derived from the predictive distributions, we further consider the mean absolute error (MAE) given by $\frac{1}{n}\sum_{i=1}^n |{x}^\text{med}_i - y_i|$, where ${x}^\text{med}_i$ denotes the median of the forecast distribution and $y_i$ the observation, and $n$ is the number of test samples, as well as the mean bias $\frac{1}{n}\sum_{i=1}^n \bar x_i - y_i$, where $\bar x_i$ is the mean value of the forecast distribution.

\subsection{Model chain}

We employ a model chain approach to obtain ensemble forecasts of PV power and apply the model chain separately to each ensemble member. Since our main aim is not to find the best possible model chain configuration, but to study the role of post-processing in this context, we directly take the model chain setup from \citet{WangEtAl2022} and implement it using code provided by the authors. 
It combines different component models \citep{Reda2004,Erbs1982,Reindl1990,king2004photovoltaic} to build the model chain, see \citet{WangEtAl2022} for details. Similar to \citet{WangEtAl2022}, we move the time stamp to the middle of the averaging period for applying the model chain, since the model chain also makes use of the time information for the estimation of the PV power output.

\subsection{Ensemble model output statistics}

As noted in the introduction, the EMOS approach proposed by \citet{gneiting2005calibrated} is one of the most widely used post-processing methods in research and operations, and will serve as a baseline method for our comparisons. \citet{PhippsEtAl2022} apply EMOS to wind speed and wind power forecasts in a similar setting, albeit using data-driven conversion models. 

The EMOS approach relies on modeling the conditional distribution of the target variable $Y$, i.e., GHI or PV power in our case, given an ensemble of predictions $x_1,...,x_m$ of the target variable via a parametric probability distribution $F_\theta$ with parameters $\theta\in\mathbb{R}^d$, i.e.,
\[
Y|x_1,...,x_m \sim F_\theta,
\]
where $\theta = g(x_1,...,x_m)$ with a link function $g$ which connects the distribution parameters with the ensemble prediction, typically via summary statistics such as the ensemble mean $\bar x = \frac{1}{m}\sum_{i=1}^m x_i$ or the ensemble variance $\text{var}(x) = \frac{1}{m-1}\sum_{i=1}^m (x_i - \bar x)^2$.
The choice of the parametric distribution $F_\theta$ plays a pivotal role in implementing EMOS models, and numerous extensions of the original normal distribution-based EMOS model of \citet{gneiting2005calibrated} 
 from temperature and surface pressure to other meteorological variables have been proposed \citep[e.g., in][]{TG2010,Lerch2013,Messner2014,Scheuerer2014,Baran2015,BaranLerch2016}. 

The EMOS models we apply here are based on censored normal distributions. For modeling GHI, several studies \citep[e.g.,][]{yang2020truncated, Yagli2020,LeGalLaSalle2020} have used forecast distributions truncated at zero, where the probability mass of the negative values is redistributed to the positive values. 
Here, we follow \citet{Schulz2021} who proposed to instead use censored forecast distributions for GHI, where the probability mass of the negative values is added to zero as a point mass.
This has the advantage that the same distribution can be used for all hours of the day, even during the night, when the GHI ensemble forecasts only contain zeros \citep{Schulz2021}. 
In contrast to \citet{Schulz2021} who applied a censored logistic distribution for GHI post-processing, we use a censored normal distribution.
For GHI, we use a normal distribution which is left-censored at zero, and for PV power we use a doubly-censored normal distribution with point masses at zero and 20  to restrict the output to the possible PV power production range in the dataset. 
The EMOS models for GHI and PV power both link the censored normal distributions (i.e., the location parameter $\mu$ and the scale parameter $\sigma$) to the corresponding ensemble forecasts of the target variable via
\begin{equation*}
\mu(x_1,...,x_m) = a + b\,\bar{x}; \hspace{1cm}
\sigma^2(x_1,...,x_m) = c + d\,\text{var}(x).
\end{equation*}
The EMOS parameters $a, b, c, d \in \mathbb{R}$ are estimated by minimizing the mean CRPS over a training dataset.
For formal definitions of the censored normal distribution and analytical formulas for computing the CRPS in closed form, see \citet{JordanEtAl2019}.

We consider two EMOS variants for both GHI and PV power. 
As a simple baseline, we train an EMOS model on data from all hours of the day by pooling all available training data into a single training dataset. This EMOS model thus applies the same correction to forecasts for every hour of the day and is not able to correct for hour-dependent errors, such as an underestimation of the PV production in the morning and an overestimation in the evening. 
To account for such diurnal variations, we further train separate EMOS models for every hour of the day, and refer to this approach as ``EMOS hourly''. 
By estimating separate sets of EMOS parameters for all hours of the day, these models thus have the advantage of being able to correct daytime-specific structures in the errors of the ensemble predictions, including systematic differences between day and night. 
A potential disadvantage of the EMOS hourly approach is that less training data is available for training the individual models. \citet{LerchBaran2017} propose alternative similarity-based estimation procedures for EMOS models that might be an interesting alternative for future studies.

\subsection{Neural network methods for post-processing}

\citet{rasp2018neural} first proposed the use of NNs for probabilistic post-processing. 
The NN approach extends the EMOS framework by replacing the link function $g$ with a NN, which connects the input predictors (e.g., summary statistics from the NWP ensemble) and the distribution parameters $\theta$, which are obtained as the output of the NN. 
The main advantages of using a NN in this context are that the NN enables the use of arbitrary input predictors, including ensemble predictions of other meteorological variables and exogeneous information, as well as the ability to flexibly model nonlinear dependencies between the inputs and the distribution parameters, which are learned in a data-driven way.
\citet{rasp2018neural} proposed a fully-connected, feed-forward NN architecture, the parameters of which are optimized using the CRPS as a loss function.
NN models for post-processing have been found to provide state-of-the-art predictive performance in many applications, and have been extended in various directions, including the use of alternative representations of the forecast distributions obtained as an output of the NN \citep{bremnes2020ensemble,SchulzLerch2022,Song2024}, or the use of more advanced NN architectures such as convolutional NNs that enable the incorporation of spatial information \citep{Scheuerer2020,Veldkamp2021,ChapmanEtAl2022,HoratLerch2024}.

We consider two different variants of NN models for post-processing, which both use the ensemble prediction of GHI (i.e., the mean and standard deviation of the ECMWF ensemble) and deterministic forecasts of 2-meter temperature and wind speed as inputs, but differ in the way they account for diurnal variations and treat the hour of the day.   
Analogous to the EMOS hourly model, we consider a NN model variant, where we train separate NN models for each hour of the day by using the corresponding subset of the training data only. 
This approach will be referred to as ``NN hourly''.
As an alternative that more efficiently uses all available data, we further consider a NN model that is trained based on all available data comprising all hours of the day.
To account for diurnal effects and differences over the different hours of the day, we provide the hour of the day information to the NN via embeddings, following a similar approach used by  \citet{rasp2018neural} to incorporate information about weather stations.
Embeddings were originally proposed in natural language processing \citep{mikolov_efficient_2013} and map categorical information to higher-dimensional latent representations in the form of vectors in $\mathbb{R}^p$. \citet{rasp2018neural} use embeddings to incorporate information about the identifier of a weather station into a NN model for post-processing, which is then jointly trained over data from all locations, but made locally adaptive by using the latent representation obtained via the embeddings as additional inputs. Here, we follow a similar strategy and learn an embedding of the hour of the day to enable the NN model to learn how to exploit diurnal patterns in the input predictors\footnote{Note that the use of embeddings ignores the temporal ordering of the hours of the day. A potential alternative for future research could be model architectures that directly use the difference of the hour of the day from the hour, where the maximum GHI value can be expected to be observed, as an input.}.
We will refer to this model as ``NN'' post-processing method\footnote{However, note that in contrast to the baseline EMOS model, despite the lack of an ``hourly'' in the model name, this approach utilizes the hour of the day information.}. 
In both NN approaches, we use a normal distribution which is left-censored at zero for GHI and a doubly-censored normal distribution with point masses in zero and 20 for PV power as in the EMOS models, and estimate the weights of the NN by optimizing the CRPS as a loss function. 

The NN architectures consist of two dense layers with 256 nodes each with ReLU activation functions, followed by one output layer with two nodes for the two distribution parameters. For the location parameter we use a linear activation, and employ a softplus activation function for the scale parameter to ensure positivity. 
For the NN with embeddings of the hour of the day, we concatenate the input forecasts with the output of the embedding layer that maps the hour of the day to a two-dimensional vector. Further, for GHI post-processing we replace the softplus activation function by a ReLU activation and add a small constant, i.e., $\text{ReLU}(x) + 10^{-3}$, to improve numerical stability during training in light of point masses at zero during the night and occasional large outliers in the deviations between predicted GHI and observed PV power during the day. 
All NN models are trained using the Adam optimizer \citep{adam2015} with a learning rate of 0.01 and early stopping, for which we use 20\% of the training data as validation data, and restore the best weights from the previous epochs. For the hourly models we use a batch size of 256 and a patience of 5 epochs during the night hours (23:00 - 5:00) 
and 30 epochs for the remaining hours to reduce the sensitivity to outliers during the day. On average, the models are trained for around 50 to 150 epochs, which is below the predefined maximum number of epochs.
For the NN models with embeddings, we use a patience of 10, a batch size of 1\,000 and train for 50 epochs (however, the early stopping always terminates the training before reaching this limit). 
For both NN approaches we standardize the predictors. It is important to note that for the hourly NN model, the standardization is done separately for each hour and exclusively based on data from the specific hour. 
To account for the stochasticity of the training process, we repeat the model training 10 times for all NN models, and use the average of the distribution parameters from the 10 runs to obtain the final predictions.

\subsection{Direct forecasting model}

As an alternative to the model chain approach with post-processing, we further consider a direct forecasting method, where we use a NN to predict PV power output directly from the available weather inputs, without the conversion via the model chain.
For this purpose, we utilize the same NN model architectures as for post-processing the PV power forecasts. 
One particular advantage of the direct forecasting models is that they do not require any intricate domain knowledge or information about the specifications of the solar plant of interest.
However, they require a training dataset of past weather predictions and corresponding PV power production to enable the model development.

\section{Results}
\label{sec:results}

We here present the results for the case study in two parts. Section \ref{sec:results-GHI} focuses on the results of post-processing the GHI forecasts using EMOS or NN approaches. 
Section \ref{sec:results-PV} then evaluates the different strategies for employing post-processing methods in the model chain approach, and provides a comparison to a direct forecasting approach without using the model chain. 

\subsection{GHI post-processing}
\label{sec:results-GHI}

\begin{figure}
\centering
\includegraphics[width=\textwidth]{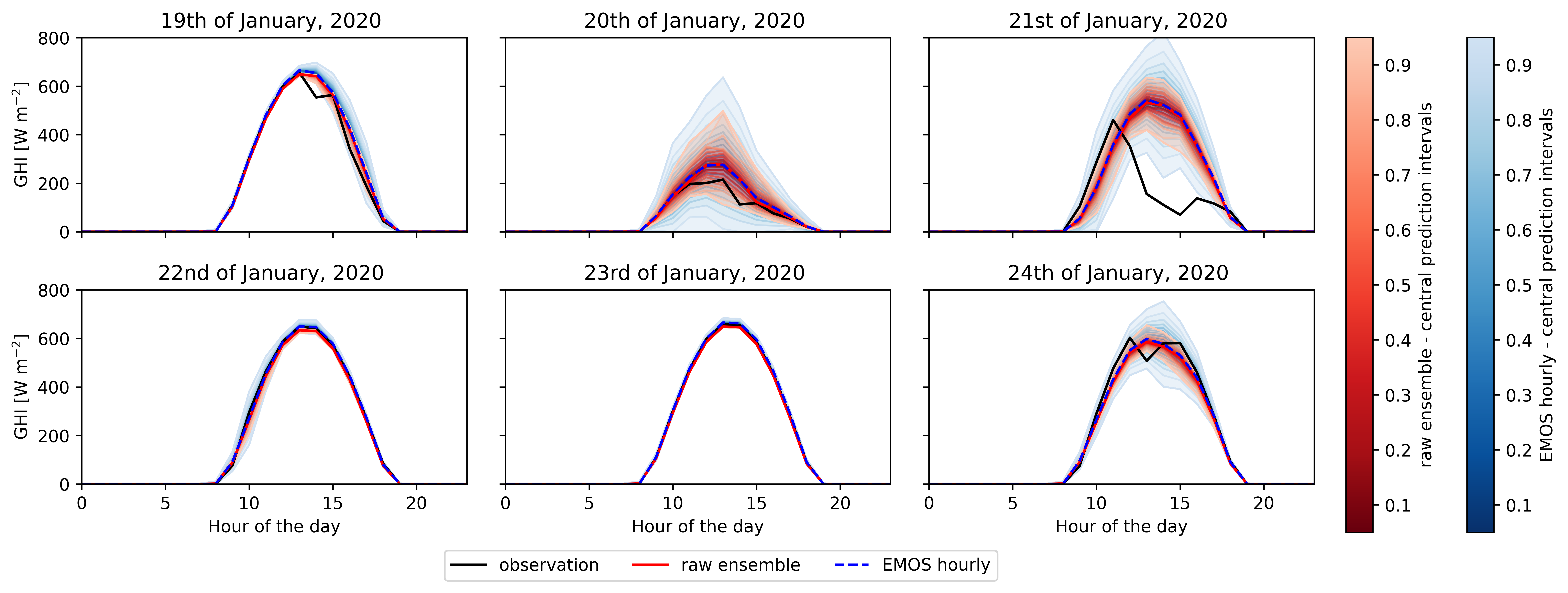}
\caption{Exemplary probabilistic GHI forecasts based on the ECMWF ensemble and the EMOS hourly post-processing approach for dates in January 2020. The colored areas indicate central prediction intervals. The lines show the ensemble median. }
\label{fig:example_GHI}
\end{figure}

Figure \ref{fig:example_GHI} shows exemplary ECMWF ensemble forecasts of GHI in comparison with observations. The observation often lies outside of the ensemble range, hence indicating that the ensemble spread is too small in the raw forecasts. 
Post-processing methods can be able to increase the spread and thereby aim to improve calibration, as exemplified by the EMOS hourly forecasts illustrated in Figure \ref{fig:example_GHI} alongside the ECMWF ensemble.
While there is considerable day-to-day variability in the forecast uncertainty, the post-processed prediction bands typically entirely encompass those of the raw forecasts.

\begin{table}
    \centering
    \caption{Mean values of various evaluation metrics for GHI forecasts from the raw ECMWF ensemble and all considered post-processing methods, averaged over all 24 hours of the day and all days in the test dataset. For the PI coverage and the PI width, also averages across daylight hours, i.e., from 6:00 - 20:00 local time are shown. PI coverage and width are computed for central prediction intervals with the nominal coverage of the raw ensemble, i.e., $(m-1)/(m+1)$ for $m=50$, which is approximately 96.1\%.}
\begin{tabular}{lcccccc}
\toprule
 & CRPS &  MAE & PI Cover. & PI Cover. & PI Width & PI Width \\
 &      &      & & daytime & & daytime \\
\midrule
ECMWF & 14.676 & 17.337 & 57.0 & 31.1 & 28.0 & \hphantom{1}44.7 \\
EMOS & 11.510 & 14.851 & 91.3 & 86.2 & 68.2 & 105.8 \\
EMOS hourly & 11.080  & \textbf{14.282} & \textbf{95.4} & \textbf{92.6} & 65.8 & 105.2 \\
NN & 11.262  & 14.725 & 94.0 & 90.4 & 66.4 & 106.2 \\
NN hourly & \textbf{11.046}  & 14.473 & 94.8 & 91.7 & 64.4 & 103.1 \\
\bottomrule
\end{tabular}
    \label{tab:GHI_crps}
\end{table}

Table \ref{tab:GHI_crps} presents evaluation scores for the ECMWF ensemble forecasts and all considered post-processing approaches. 
As expected, post-processing shows substantial improvements over the raw ensemble forecasts, for example, of up to 25\%  in terms of the mean CRPS and around 18\% in terms of the MAE. 
Clear improvements are also achieved in terms of coverage of central prediction intervals. While the raw ECMWF ensemble predictions provide the sharpest forecasts with the shortest prediction intervals, they lack calibration since on average, only 31.1\% of the daylight observations lie within the 96.1\% prediction interval. By contrast, all post-processing methods provide better calibrated forecasts and achieve coverages close to the nominal level, with substantially wider prediction intervals.

All post-processing methods perform more or less equally well, with the hourly approaches achieving slightly better scores than the methods trained on data from the entire day.
The NN approaches outperform the EMOS variants slightly, even though the relative differences between NN hourly and EMOS hourly are less than 1\% in terms of the mean CRPS. Similar conclusions apply to the other evaluation metrics.
In contrast to other applications of NN methods for post-processing, these improvements are on a much smaller scale. 
However, they are in line with previous findings which indicate that the main advantage of using NN methods for post-processing is the efficient use of additional input information \citep{rasp2018neural}.
The additional inputs available to the NN models here likely do not provide substantial predictive information about GHI, and the main advantage of the NN approaches over the EMOS models thus might be given by the potential to learn non-linear link functions, see also \citet{EUPPBench} for related results.

\begin{figure}
\centering
\includegraphics[width=\textwidth]{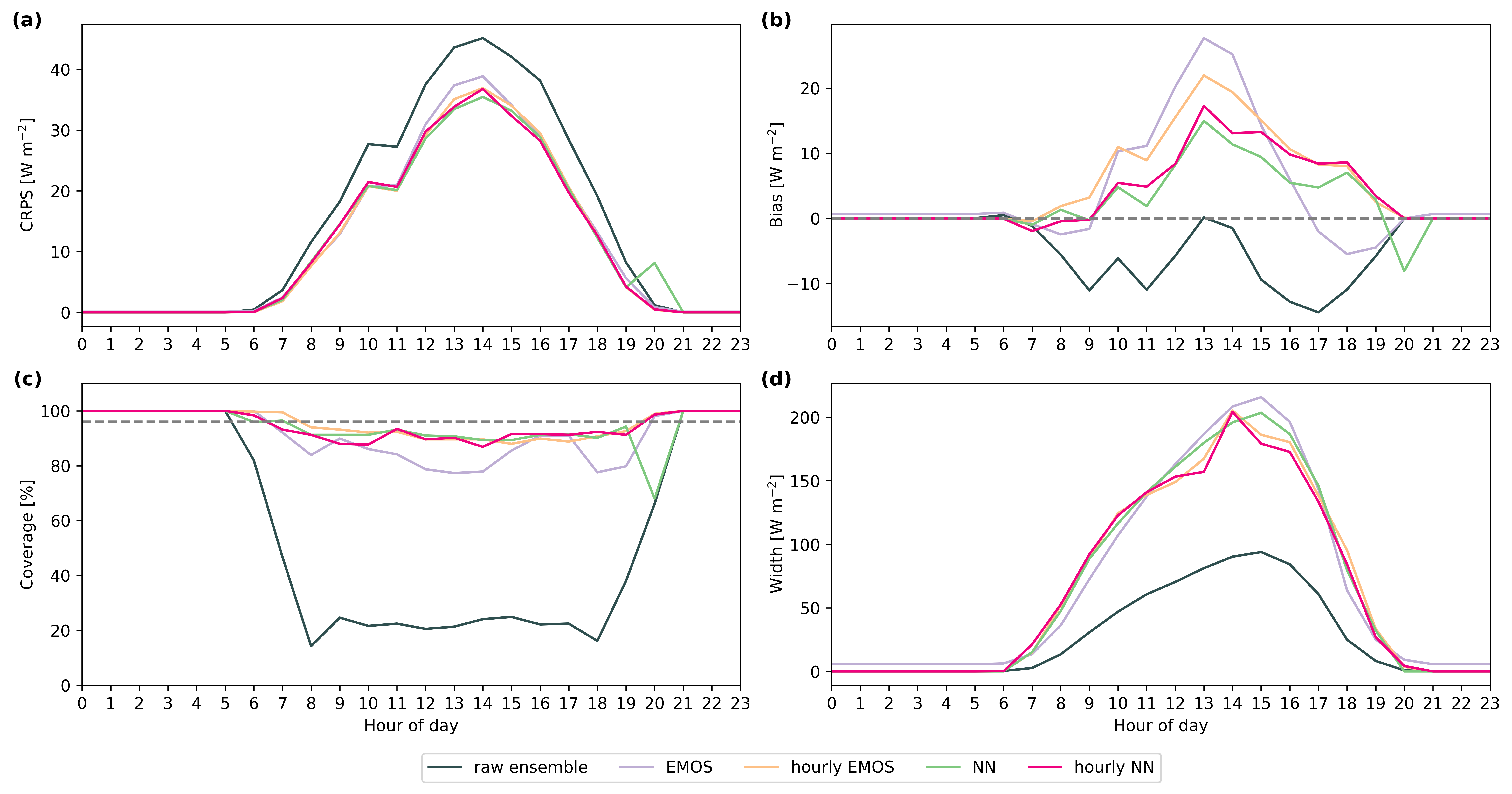}
\caption{Hourly values of the CRPS (a), the bias (b) of the mean forecast, as well as the coverage (c) and width (d) of 96.1\% prediction intervals for the GHI forecasts. All values are averaged over the test dataset.}
\label{fig:GHI_scores_hourly}
\end{figure}

\begin{figure}
\centering
\includegraphics[width=\textwidth]{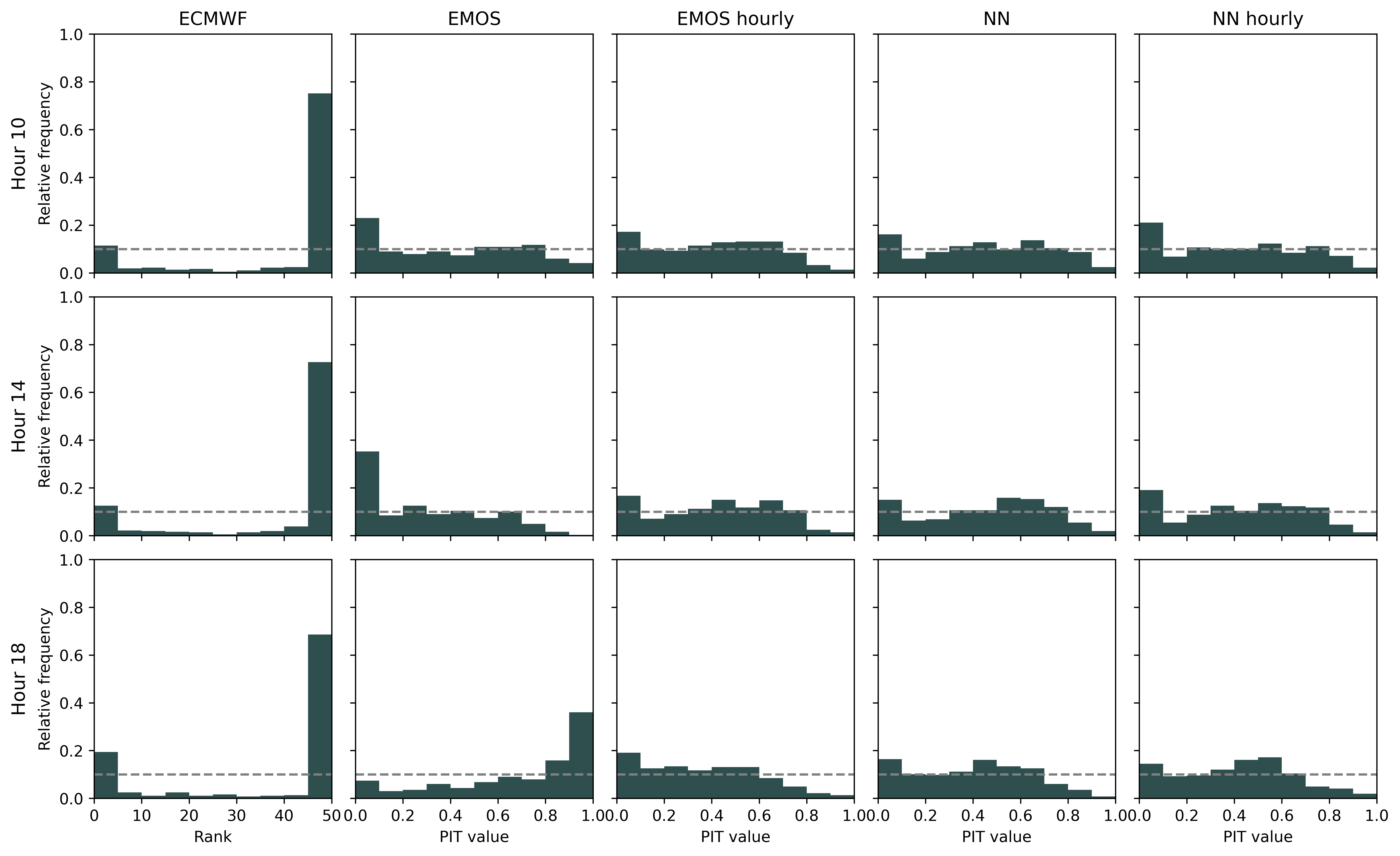}
\caption{Verification rank histogram of the ECMWF ensemble forecasts and PIT histograms for all considered post-processing methods for GHI for hours 10, 14, and 18.}
\label{fig:GHI_PIT}
\end{figure}

Figure \ref{fig:GHI_scores_hourly} shows the evolution of various evaluation metrics over the course of the day.  
As expected, all evaluation metrics show a strong dependence on the time of the day, and the CRPS, bias and width of the prediction intervals are almost zero for all methods during nighttime. During nighttime, all forecasts coincide at zero, which prevents a proper computation of coverage and PI width, since the coverage equals one for all levels, and the corresponding PI width is always zero.
The CRPS curves of all forecasts show a maximum at around hour 14, and the curves for almost all post-processing methods lie below the CRPS curve of the raw ensemble for almost all hours of the day. The only notable exception is the NN model at hour 20, which likely corresponds to numerical stability issues in the parameter estimation as this outlier is also present in the bias and coverage curves.\footnote{A more detailed investigation suggests that this might be due to notable violations from the distributional assumptions, indicated by heavily skewed histograms of the GHI observations at this hour with most observations at 0, but a long tail with values up to 50. While such distributions might be challenging to model with a censored normal distribution in general, the estimation of the NN models seems to show particular difficulties in converging to reasonable parameter estimates.
Non-parametric methods such as Bernstein Quantile Networks \citep{bremnes2020ensemble, GneitingEtAl2023} or quantile regression \citep{Song2024} could provide a possible remedy by also allowing for non-symmetric and non-normal distributions. For example, \citet{GneitingEtAl2023} show examples comparing NNs learning distributional parameters and BQN methods and report better CRPS score for the BQN method than for the parametric NN approaches for GHI due to the enhanced flexibility of the BQN method. 
}
Interestingly, the CRPS curves for EMOS, i.e., the only post-processing model that does not use information about the hours of the day, closely follows the CRPS curves of the other post-processing models except for hours around midday, where the model shows a larger CRPS.
The MAE curves look almost identical to those of the CRPS and thus are omitted here.
All post-processing models show a slightly positive bias with a maximum around midday. Since the raw ensemble shows a negative bias throughout the day, the sign (but not the magnitude) of the bias changes due to post-processing. 
The NN approaches show smaller biases during most of the day, while the bias structure of the EMOS model reveals that the same bias correction is applied to all hours of the day and the model is unable to learn hour-dependent error characteristics.
However, given the magnitude of the bias of all methods, the improvements in terms of the CRPS largely stem from an improved calibration of the forecasts. 
This also becomes evident from the coverage and width of the prediction intervals from the different methods. Not surprisingly, the raw ensemble forecasts produce the shortest prediction intervals and thus the sharpest forecasts throughout the day, however, they fail to achieve a coverage close to the nominal value. All post-processing methods yield substantially wider prediction intervals and a better coverage. Again, the EMOS model shows a slightly worse performance than all other methods that use predictive information about the hour of the day. 
In terms of the prediction interval width, it is interesting to note that the hourly EMOS and NN methods yield slightly sharper forecasts than their corresponding more general counterparts.
A closer inspection of the performance of the EMOS model during nighttime indicates that neither the bias nor the prediction interval width are exactly zero. This approach thus fails to appropriately model the GHI values during nighttimes as a point mass in zero.

To further assess the calibration of the probabilistic forecasts, Figure \ref{fig:GHI_PIT} shows verification rank and PIT histograms of all approaches for selected hours of the day.
The raw ECMWF ensemble forecasts of GHI are clearly underdispersive and lack calibration, as indicated by the U-shaped verification rank histograms.
The PIT histograms of all post-processed forecasts are notably closer to the desired uniform distribution, and thus indicate that these forecasts are better calibrated. 
The EMOS hourly and the two NN approaches show the best calibration.
By contrast, the EMOS model jointly estimated over all hours of the day produces less well calibrated forecasts and a clear bias in the form of an overestimation of the GHI values at hour 10, and an underestimation at hour 18, respectively.

\subsection{PV power forecasts}
\label{sec:results-PV}

Here, we first present the results for the different strategies of applying post-processing in a model chain approach, and then compare to a direct forecasting model.

\subsubsection{Post-processing in the model chain approach}

\begin{table}
    \centering
        \caption{Mean CRPS values for probabilistic PV power forecasts obtained from the considered post-processing strategies using different post-processing methods. Note that in the \pppp strategy, the same post-processing method is applied in both steps. All CRPS values are averaged over all hours of the day. The model chain approach without any post-processing (\rawraw\!\!) achieves a CRPS of around 0.689.}
\begin{tabular}{lcccc}
\toprule
Strategy  & EMOS & EMOS hourly & NN  & NN hourly \\
\midrule
GHI$^\text{pp}$-PV$^\text{raw}$ & 0.676 & 0.651 & 0.639 & 0.644 \\
GHI$^\text{raw}$-PV$^\text{pp}$  & 0.564 & 0.305 & \textbf{0.294} & 0.309 \\
GHI$^\text{pp}$-PV$^\text{pp}$ & 0.573 & 0.306 & \textbf{0.294} & 0.308 \\
\bottomrule
\end{tabular}
    \label{tab:PV_crps}
\end{table}

\begin{figure}
\centering
\includegraphics[width=\textwidth]{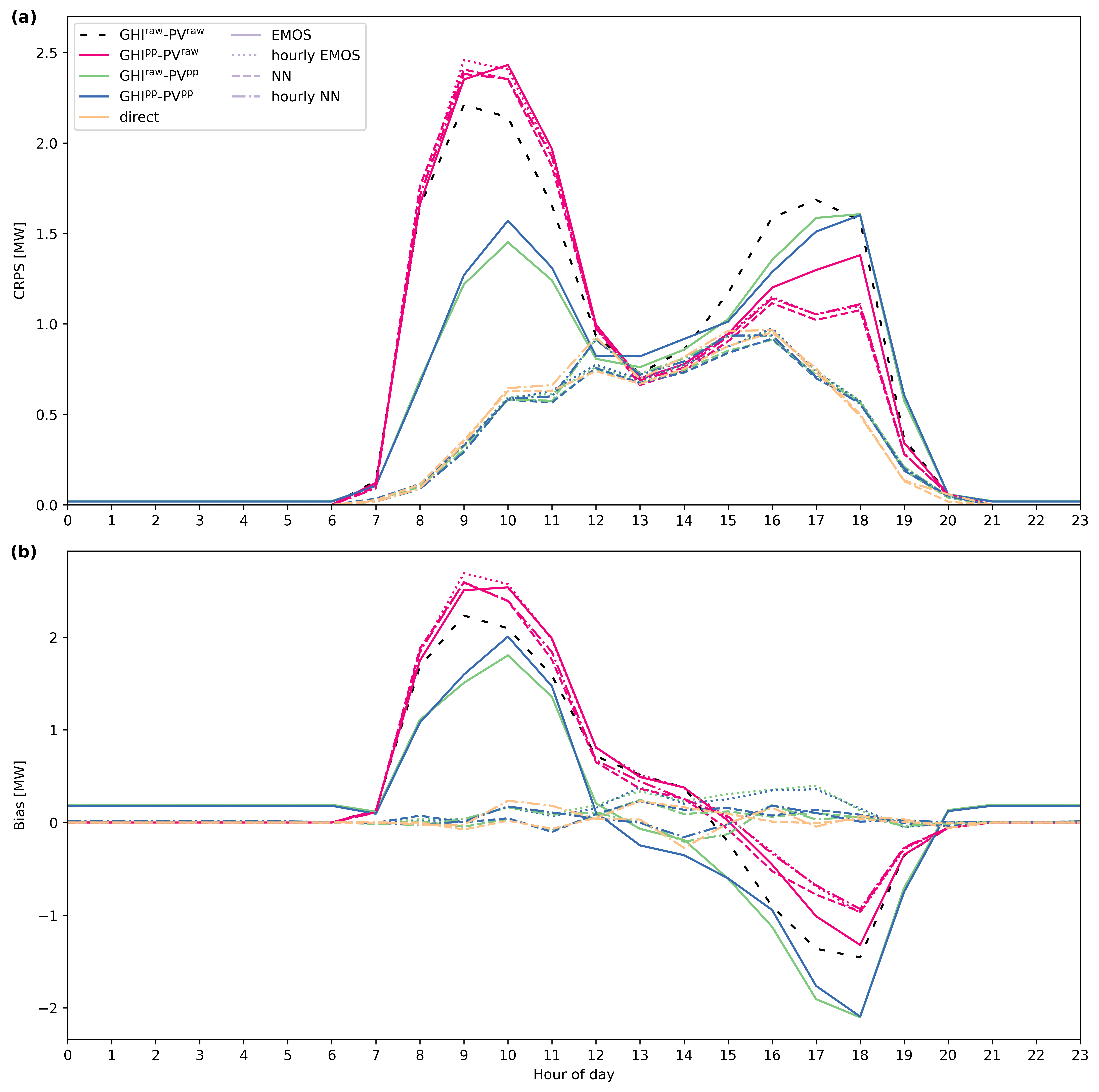}
\caption{Hourly values of the mean CRPS (a) and the bias (b) of the mean forecast for the considered strategies and post-processing methods, as well as the direct forecasts. Note that the strategies are indicated by line type, and the post-processing methods by the color of the corresponding lines.}
\label{fig:PV_scores_hourly}
\end{figure}

Table \ref{tab:PV_crps} shows mean CRPS values for all combinations of strategies for applying post-processing and post-processing methods. 
All combinations improve the PV power forecasts compared to using the model chain approach without any post-processing, but the magnitude of the improvements differ substantially across methods and strategies.
Applying post-processing only to the GHI forecasts in the \ppraw strategy leads to the smallest improvements of at most around 7\% in terms of the mean CRPS. 
For all post-processing methods, applying post-processing to the PV power forecasts obtained from the model chain appears to be the most crucial step, as the \rawpp and \pppp strategies achieve almost identical mean CRPS values for all post-processing methods, and improvements over the raw model chain forecasts of up to around 57\%.
In terms of the different post-processing methods, the EMOS model jointly estimated for all hours of the day performs substantially worse than all others. 
The best overall CRPS values for all strategies are achieved by the NN model which uses the hour of the day information via embeddings.
In contrast to the GHI forecasting task, where this model performed worse than the two hourly approaches, it thus might be more beneficial for PV power forecasting to utilize a NN with an increased training sample size.
The two hourly models, EMOS hourly and NN hourly, achieve very similar CRPS values, with slightly better scores for the simpler EMOS hourly model. As discussed in the results for the GHI predictions, the benefits of using a NN approach here might again be limited by the information content of the additional predictors available to the NNs. Further, the EMOS model is notably simpler to tune, with fewer hyperparameters and optimization settings that need to be chosen.

To assess the diurnal variability of the forecast errors, Figure \ref{fig:PV_scores_hourly} shows hourly values of the mean CRPS and bias for all combinations of strategies and post-processing methods. Note that the figure also contains results for the direct forecasts, which we will discuss below. 
Both the model chain approach based on the raw ECMWF ensemble predictions without any post-processing, as well as the \ppraw strategy, independent of the post-processing method, show substantially larger forecast errors with a pronounced diurnal cycle. The largest forecast errors occur during the early morning and late afternoon hours, whereas the CRPS during midday is comparable to that of the other strategies and post-processing methods.
A similar behavior, albeit with smaller errors in the morning, but larger errors in the afternoon, can be observed for the EMOS approach estimated jointly over all hours of the day in the two remaining strategies (\rawpp and \pppp\!\!\!). 
The main explanation for these observations likely is the behavior of the bias, which is notably larger than it was for the GHI forecasts when compared to the magnitude of the CRPS. 
The less well performing strategies and methods (i.e., the raw model chain without post-processing, the \ppraw strategy, and all non-hourly EMOS variants) show substantially larger biases that change the sign from positive (i.e., overestimation) in the morning to negative (i.e., underestimation) in the afternoon.
Not surprisingly, the EMOS model which is jointly estimated over all hours of the day is not able to account for these daytime-specific variations in the bias.
All other combinations where post-processing is applied to the PV power forecasts obtained as output of the conversion model and information about the hour of the day enters the model achieve notably smaller biases and lower CRPS values during the whole day. The relative differences between these approaches are only minor, with slightly increased biases of the EMOS hourly forecasts during the afternoon.

\begin{figure}
\centering
\includegraphics[width=\textwidth]{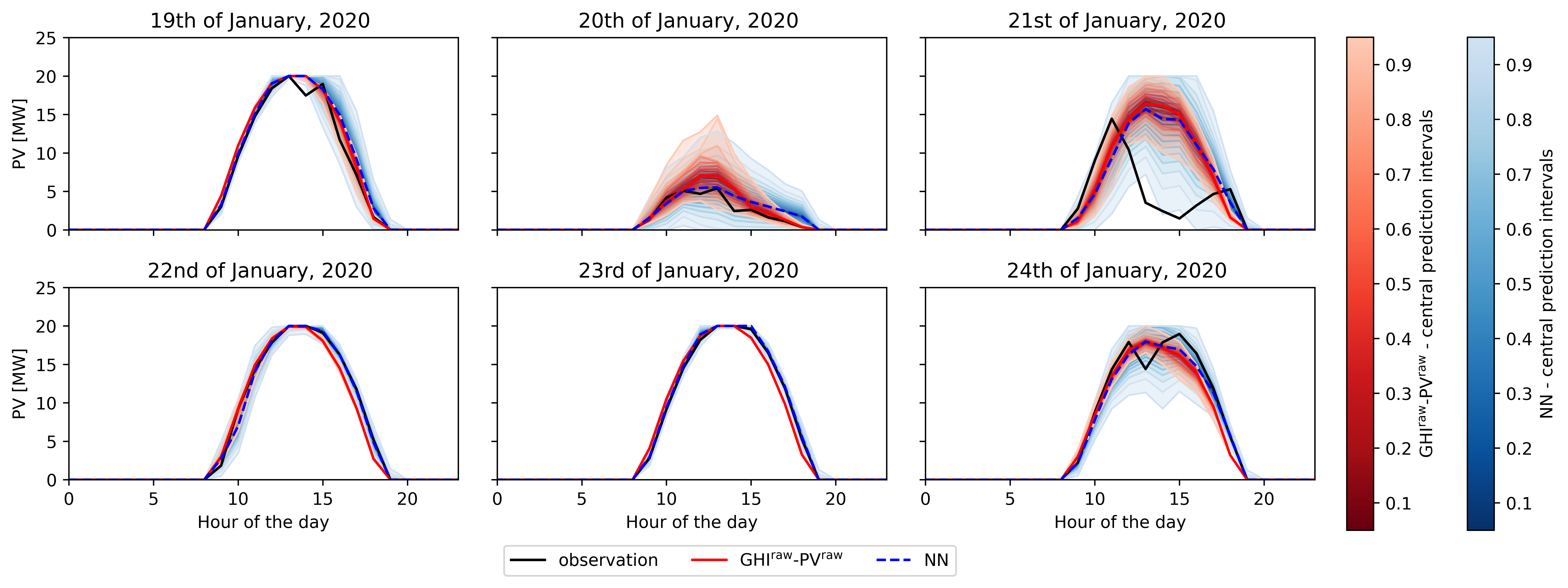}
\caption{Exemplary probabilistic PV forecasts for dates in January 2020 for \rawraw and \pppp with the NN jointly estimated for all hours of the day. The colored areas indicate central prediction intervals. The lines show the ensemble median. }
\label{fig:PV_example}
\end{figure}

Due to the large number of combinations of strategies and methods, the following graphical illustrations and corresponding discussions focus on results for the best-performing strategy (\pppp\!\!) and/or method (NN).
Figure \ref{fig:PV_example} shows exemplary probabilistic PV power forecasts and corresponding observations for the model chain without any post-processing and the \pppp strategy with the NN model for post-processing for the same days as the GHI forecasts in Figure \ref{fig:example_GHI}. 
A large variability in the forecast uncertainty can be observed over the different days, which seems to be directly connected to the forecast uncertainty of the corresponding GHI predictions. 
As for GHI, post-processing here substantially increases the width of the prediction intervals.

\begin{figure}
\centering
\includegraphics[width=\textwidth]{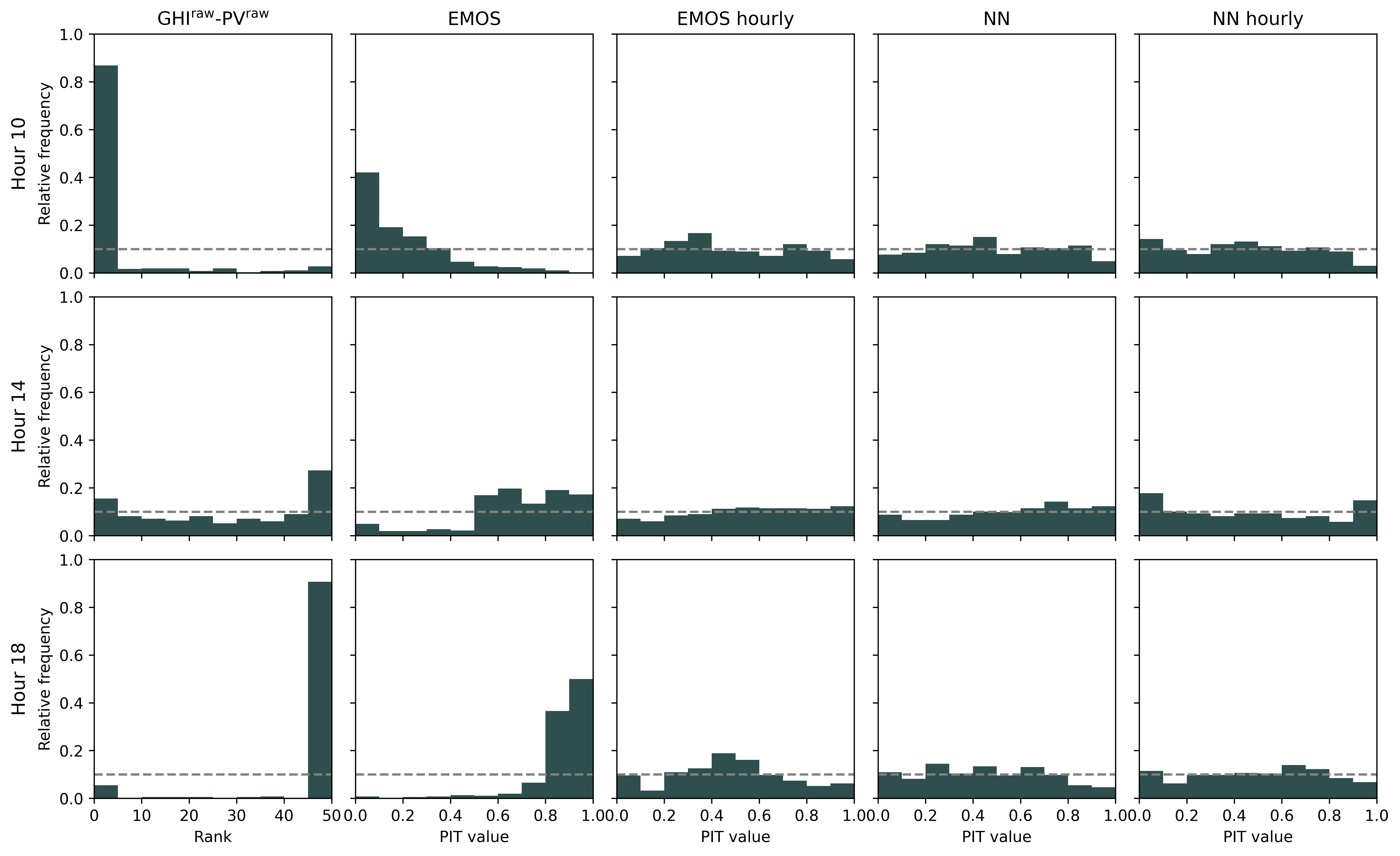}
\caption{Verification rank histogram and PIT histograms for the raw forecasts (\rawraw\nolinebreak\!\!) and all considered post-processing methods for PV power output within the \pppp approach for hours 10, 14, and 18.}
\label{fig:PV_PIT}
\end{figure}

Figure \ref{fig:PV_PIT} shows verification rank and PIT histograms for the \rawraw and \pppp strategies and all considered post-processing methods. In light of the biases observed in Figure~\ref{fig:PV_scores_hourly}, it is not surprising that neither the model chain without any post-processing nor the \pppp strategy with EMOS post-processing show calibrated forecasts, but clearly notable biases, in particular during hours 10 and 18. 
All other post-processing methods yield substantially better calibrated forecasts with PIT histograms much closer to uniformity.

\subsubsection{Direct forecasting model}

\begin{table}
    \centering
    \caption{Mean CRPS values for probabilistic PV power forecasts obtained from the \pppp post-processing strategies using NN methods in comparison to NN-based direct forecasting methods that do not utilize a conversion from GHI to PV power via the model chain. Scores are averaged over all 24 hours of the day and all days in the test dataset. For the PI coverage and the PI width, also averages across daylight hours, i.e., from 6:00-20:00 local time are shown. PI coverage and width are computed for PIs with the nominal coverage of the raw ensemble, i.e., $(m - 1)/(m + 1)$ for $m = 50$, which is approximately 96.1\%.
}
\begin{tabular}{lcccccc}
\toprule
 & CRPS & MAE & PI Cover. & PI Cover. & PI Width & PI Width\\
  &      &      & & daytime & & daytime \\
\midrule
\rawraw & 0.698 & 0.768 & 65.4 & 44.7 & 0.686 & 1.098 \\
\midrule 
\pppp NN & \textbf{0.294} & \textbf{0.397} & 97.5 & \textbf{96.0} & 1.808 & 2.894 \\
\pppp NN hourly & 0.308 & 0.413 & 97.8 & 96.4 & 2.773 & 4.317 \\
\midrule 
Direct NN & 0.298 & 0.409 & 97.9 & 96.7 & 1.841 & 2.945 \\
Direct NN hourly & 0.314 & 0.422 & \textbf{96.1} & 93.8 & 2.535 & 4.056 \\
\bottomrule
\end{tabular}

    \label{tab:PV_direct}
\end{table}

Table \ref{tab:PV_direct} shows various evaluation metrics for the NN-based direct forecasting methods for PV power that do not utilize a conversion from GHI to PV power via the model chain.
Both the direct NN model jointly estimated for all hours via embeddings and the hourly direct NN model show substantially better CRPS values than the model chain approach without any post-processing.
The mean CRPS values achieved by the direct forecast models are comparable to those of their \pppp counterparts, albeit slightly worse.
Revisiting Figure \ref{fig:PV_scores_hourly}, we note that the diurnal pattern of the direct models is almost identical to that of the corresponding \pppp model.  The two approaches are also very similar in terms of coverage and PI width. 
For the daylight hours, all NN approaches achieve a coverage which is very close to the nominal coverage of 96.1\%. However, the hourly models have prediction intervals that are up to 50\% wider than the models estimated on data from all hours of the day. An investigation of the width of the prediction intervals for every hour of the day (not shown) reveals that the PIs for 13:00 and 14:00 local time are extremely wide, reaching up to 18 MW. As discussed above, we assume that for these hours, the assumption of a doubly-censored normal distribution is violated since the distribution of the observed PV power is heavily skewed. 

\section{Discussion and conclusions}
\label{sec:conclusions}

We have systematically compared different strategies for employing post-processing to improve probabilistic PV power forecasts within a model chain approach, where weather predictions are converted to PV power via a cascade of physics-based models. 
In a case study for a solar plant in the U.S.\ based on data from \citet{WangEtAl2022}, we develop statistical and machine learning methods for post-processing GHI and PV power forecasts.
We find that post-processing leads to substantial improvements when applied to the PV power forecasts that are obtained as the output of the model chain, in line with findings from \citet{PhippsEtAl2022} in the context of wind energy prediction.
Whether or not the GHI forecasts are post-processed before using them as input to the model chain plays an almost negligible role for most post-processing methods.

In terms of the performance of different post-processing approaches, the use of the hour of the day is of central importance when building a model, either via utilizing separate models for each hour of the day, or by including the temporal information as input to a NN model, in our case via embeddings.
Comparing classical EMOS and modern NN-based models for post-processing, we find that in contrast to various recent studies on other weather variables, the use of NNs here only leads to minor improvements. 
A likely explanation of this finding is that the additional input information that was available to the NN models carries too little predictive information to be effectively utilized, since we restricted our attention to those meteorological variables that also served as an input to the model chain (i.e., deterministic temperature and wind speed predictions).

We have further proposed a NN model for directly predicting PV power output from the weather information without using a model chain for the intermediate conversion of GHI to PV power. 
This direct forecasting model showed almost competitive forecast performance with the best combination of post-processing strategy and method in the model chain setting, but comes with the advantage of being applicable without requiring specific knowledge about the individual solar plant's design and technical specifications. However, estimating a direct forecasting model of course requires past weather predictions and PV power observations as training data, which is not necessary for a model chain approach, at least if no post-processing is applied.

Our study provides several avenues for further model development and analysis.
Perhaps most importantly, it only constitutes a first step towards a more systematic analysis and comparison of the different strategies and post-processing methods, since we only used data from a single solar plant and a single model chain approach. 
Repeating the analysis on a more comprehensive dataset with multiple locations and potentially an ensemble of model chains \citep{Mayer2022} would not only provide a more comprehensive comparison, but would also allow for addressing interesting methodological questions, e.g., how to effectively develop NN model architectures for multiple sites, or how to post-process multi-model ensembles in this setting.

Another natural starting point for future research are ways to further improve our NN models for post-processing and direct PV power prediction. Instead of learning distribution parameters with the NNs, non-parametric methods would allow for non-symmetric and non-normal distributions of the target variable and could provide a better fit for the heavily skewed distributions of GHI and PV power during midday, early morning or late evening hours, see \citet{bremnes2020ensemble}, \citet{SchulzLerch2022}, and \citet{GneitingEtAl2023} for examples.
Further, using deterministic predictions of more weather variables from the data available in \citet{WangEtAl2022} as input to the NN models might lead to improvements. 
Note that we only post-processed the GHI predictions, but not the other inputs to the model chain and the NN models. If observations for those variables were available, post-processed forecasts might further improve the performance of the GHI to PV power conversion via the model chain. In all of these developments, it would furthermore be interesting to consider additional aspects of forecast quality beyond statistical evaluation metrics, such as economic aspects \citep{vanderMeerEtAl2018,Gneiting2023}.

Finally, over the past few years, there have been rapid developments in AI-based data-driven weather models such as Pangu-Weather \citep{BiEtAl2023} or GraphCast \citep{LamEtAl2023}.
Since those models have been demonstrated to outperform classical physics-based NWP models on a variety of prediction tasks, investigating whether they could also replace NWP models as inputs for solar energy prediction might be an interesting question for future research. The use of post-processing methods investigated in our article might play an important role in this context, since these AI weather models likely exhibit different systematic error characteristics, and model chains adapted to physics-based inputs thus might not work as well. Further, many currently available models do not provide relevant outputs like GHI and are limited to deterministic predictions only. They thus require additional steps for quantifying forecast uncertainties, which are particularly relevant in applications such as solar energy prediction \citep{buelte_etal_2024}.

\section*{Acknowledgments}

The research leading to these results has been done within the Young Investigator Group ``Artificial Intelligence for Probabilistic Weather Forecasting'' funded by the Vector Stiftung. In addition, this project has received funding from the Federal Ministry of Education and Research (BMBF) and the Baden-Württemberg Ministry of Science as part of the Excellence Strategy of the German Federal
and State Governments. We thank Peter Knippertz, Wenting Wang and Dazhi Yang for helpful comments and discussions.

\bibliographystyle{myims2}
\bibliography{bibliography}

\end{document}